\documentclass[]{acmtrans2m}

\usepackage{graphicx}
\usepackage{booktabs}
\usepackage{subfig}

\newcommand\eg{e.\,g.,\ } %......exempli gratia (example given)
\title{Performance analysis of Xen virtual machines in real-world scenarios} 
\author{
ADRIAN HEISSLER\\University of Applied Sciences Technikum Wien\\
}

\begin{abstract}
This paper presents results of the performance benchmarks of the Open Source
hypervisor Xen.  The study focuses on the network related performance as well
as on the application related performance of multiple virtual machines that were
running on the same Xen hypervisor.  The comparison was carried out using a 
self-developed benchmark suite that consists of easily available Open Source
tools. The goal is to measure the performance of the hypervisor in typical
real-world application scenarios when used for ``mass virtual hosting'', such as  
hosting solutions of so called virtual private servers for small-to-medium sized
businesses environments.  The results of the benchmarks show, that the tested
Xen setup offers good performance with respect to network traffic stress tests, 
but only 75\% of the performance of the non-virtualized reference environment. 
This application performance score decreases as more virtual machines are 
running simultaneously.
\end{abstract}




\begin{document}

\setcounter{page}{1}

\maketitle

% -----------------------------------------------------------------------------

\section{Introduction}

System virtualization has become an important tool in the information technology
community.  It is useful in many scenarios, \eg server consolidation or
rapid deployment of new virtual servers.  To better utilize the hardware of a
physical machine, the main goal in real-world scenarios often is to run as many
virtual machines as possible on the same physical host.  

In this paper, the networking performance of virtual machines with respect to
the metrics latency and throughput is measured and analysed.  Furthermore, the
performance of application programs that run inside virtual machines is
benchmarked and analysed.  The study deals with virtual machines that are
created by the popular Open Source hypervisor Xen.  The chosen application 
benchmark scenarios, reflect typical application programs that are used by 
small-to-medium businesses and that may be virtualized.  These scenarios include
a web server, database server, and file server.

Many of the commonly used benchmarks suites that are used to study server
consolidation scenarios are only available as commercial products.  This paper 
proposes a benchmark suite that consists of easily available and well known Open
Source programs.

The benchmarks that have been carried out, also try to honor a typical setup
scenario at many sites, such as hosting providers of virtual private servers:
For maintenance reasons, the disk images of the virtual machines are often
stored on a network shared storage that is accessed via iSCSI or NFS, \eg to
enable live migration of virtual machines between physical hosts.

% -----------------------------------------------------------------------------

\section{Related work}

The measurement of the performance of hypervisors like Xen has been subject to
many studies, including \citeN {barham2003}, \cite{clark2004}, 
\citeN{deshane2008}, \citeN{apparao2006}, \citeN{cherkasova2005}, 
\citeN{matthews2007}, \citeN{xu2008}, \cite{tanaka2009}.

In addition to the benchmark suite proposed in this paper, well known benchmark
suites are IBM Virtualization Grand Slam benchmark \cite{ibm2004}, vConsolidate 
\cite{casazza2006,apparao2008}, and VMmark \cite{vmware2006}.

% -----------------------------------------------------------------------------

\section{Setup}

The test platform is a HP ProLiant DL380 G5 with 2 Quad Core Intel Xeon
processors (2.66 GHz, 4 MB cache size), 16 GB RAM, two PCI-E Dual Port
Multifunction Gigabit network controllers, and a HP Smart Array controller (two 
140 GB SAS disks are configured in a RAID 1). The setup consists of CentOS 5.4 
(i386) on the host machine.  For the Xen hypervisor tests, kernel 
2.6.18-164.6.1.el5xen and Xen 3.0.3\footnote{ Since the distribution release 
5.2, the CentOS Xen 3.0.3 package includes in fact selected backports of Xen 
3.1.2.} are used.  All the software packages are installed from the official 
CentOS repositories.  The SMP Credit Scheduler has been used as the hypervisor 
scheduler throughout all benchmarks.  The virtual machines are run in
para-virtualized mode.

The disk images of the virtual machines are stored on a NFS share on top
of a NetApp FAS3140 cluster.  The host machine is connected to the NetAPP filer
via a Gigabit Ethernet link, which is dedicated to serve only connections
between the host and the NetApp filer.

The benchmark suite is twofold:  the first part deals with network performance
of the virtual machines, the second part details on the performance of typical
application scenarios that may be run virtualized.

The maximum network data transfer rate of each virtual machine has been limited
to 50 Mbit/s during all the tests.

% -----------------------------------------------------------------------------

\section{Evaluation of network data rate and latency of virtual machines}

The goal of this experiment is to measure the performance of the virtual
machines' virtual interfaces (VIF) with respect to two metrics --- throughput 
and latency.  For this experiment, all virtual machines are set up with 1
virtual CPU (vCPU), 512 MB RAM, 1 GB swap space and a 18 GB virtual hard disk. 
CentOS 5.4 with kernel 2.6.18-164.6.1.el5xen has been installed from the
official CentOS repository on the virtual machine.

To evaluate the TCP performance of a virtual machine, \textit{iperf}
\cite{gates2008} version 2.0.4 has been used to measure the maximum achievable 
network throughput (goodput) between the virtual machines and external physical 
hosts. For network latency the \textit{ping} utility is used to measure the 
packet round trip time (RTT) between the virtual machines and the external 
physical hosts.

Several test runs have been executed, starting with only a single virtual
machine.  Then the number of virtual machines that were each concurrently 
running the tests, is increased.  All virtual machines were running on the same
host.

Each virtual machine has one of three external physical hosts as a ``partner''
during each test run.  The external physical hosts are connected via a Gigabit 
link to the same switch as the Xen host.

A test run for a certain virtual machine consists of two iperf tests and a ping
test. Each iperf test was run for 60 s.  One iperf run tests the sending 
capabilities, the second iperf run tests the receiving capabilities of the
virtual machine. For the iperf tests, the TCP window size has been set to the
default value of 16 KB on the sender and receiver.  The ping utility is run for 
10 s during every first iperf run.  It sends 64-byte ICMP messages to the remote
host.

\begin{figure}
\centering
\includegraphics[scale=0.9]{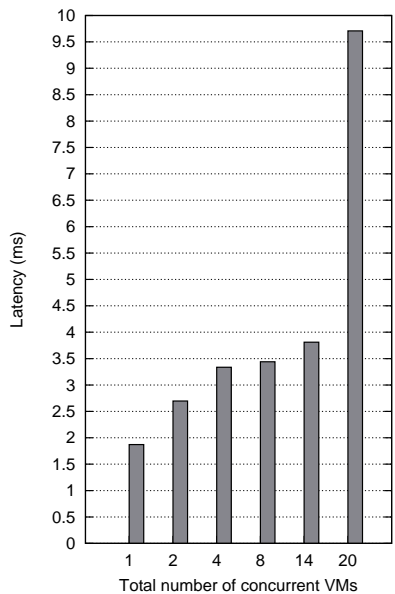}
\caption{Aggregated average RTT of the virtual machines during the network
throughput tests.}
\label{fig:vserver-node-ping}
\end{figure}

\begin{figure}
\centering
\includegraphics[scale=0.9]{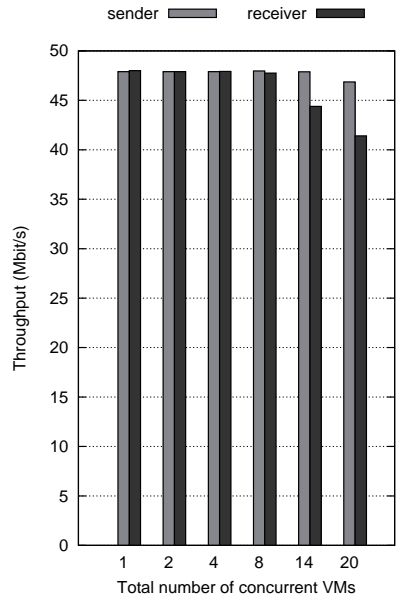}
\caption{Aggregated average TCP throughput of the virtual machines.}
\label{fig:vserver-node-iperf}
\end{figure}

According to \citeN{raj2007}, there are two possibilities how virtualization can 
introduce latency.  Firstly, a packet must be classified to which VIF it belongs
to. Secondly, the guest domain that owns this VIF has to be be notified. 

When measuring the latency for virtual machines with no network load introduced,
the avarage RTT was around 0.232 ms.  As shown in figure
\ref{fig:vserver-node-ping}, the RTT increases as the number of virtual machines
increases that were each performing the network throughput test.  Only with 20
virtual machines concurrently performing the iperf test, the average latency
significantly increases to almost 10 ms.  The increase of latency is mainly 
produced by CPU contention, as well as by increased context switches and 
interrupt servicing in the Xen driver domain (Domain-0) on the host 
\cite{apparao2006}.

Figure \ref{fig:vserver-node-iperf} shows the iperf results for both, the sender
and receiver test.  Both streams can sustain almost 50 Mbit/s during the entire
tests up to 14 virtual machines concurrently introducing network load.  With 14
and 20 virtual machines the throughput decreases, particularly for the receiver
tests.

The gap between achieved throughput and theoretical data transfer rate is mostly
due to the fact that TCP's features such as flow control mechanisms, are 
limiting the throughput.  Thus, TCP is often not able to fully utilize the
available network data rate.

% -----------------------------------------------------------------------------

\section{Application performance evaluation of virtual machines}

In order to evaluate the performance of the virtual machines a methodology 
partly inspired by benchmarks proposed by \citeN{ibm2004}, 
\citeN{casazza2006}, and \citeN{vmware2006} has been used.

The benchmark uses three application environments, each representing a different
application that would be typically run on a virtual machine.  The benchmark 
consists of the following three application environments:

\begin{itemize}
  \item Apache web server.  \textit{Siege} \cite{fulmer2009} version 2.69 is 
  used for benchmarking the Apache HTTP server.  This benchmark is run from an 
  external physical machine (this machine is interconnected to the Xen host
  via a single switch). A workload of 25 concurrent users accessing a 65 KB
  static HTML site is simulated.
  \item PostgreSQL database server.  The \textit{pgbench} \cite{postgresql2009} 
  program is used here for the benchmarks.  Pgbench is a simple program for 
  running benchmark tests on a PostgreSQL database.  This benchmark is run 
  locally on the tested virtual machine.
  \item Samba file server.  The \textit{dbench} \cite{tridgell2008} application 
  version 4.0 is used to simulate file system load by performing all the same 
  I/O calls that a server message block (SMB) server in Samba would produce when
  confronted with a NetBench run. This benchmark is run locally on the tested 
  virtual machine with 48 simulated clients.
\end{itemize}

Each of these applications has been installed on three different virtual 
machines with different Linux distributions as operating systems.  Again,
only software packages from the official distribution repositories have
been installed on the virtual machines.  These virtual machines are configured
with 1 vCPU, 2 GB RAM, 4 GB swap space, and 72 GB virtual hard disk.

Furthermore, an idle server has been set up with 1 vCPU, 512 MB RAM, 1 GB
swap, and 18 GB virtual hard disk.  Although idle, this system still place
resource demands upon the virtualization layer and can impact the performance of
the other virtual machines.

All four virtual machines have been started on the same physical host,  no other
virtual machines were running on this host during the tests.

This host machine itself also acts as a reference system.  For its own test
series it has been configured with 1 CPU and 2 GB RAM.

Table \ref{tab:infrastructure_performance} summarizes the different workload 
profiles used throughout the measurements.

\begin{table}
\centering
\begin{tabular}{lcccc}
\toprule
Resource & Web server & Database server & File server & Idle \\
\midrule
CPUs (\#) & 1 & 1 & 1 & 1 \\
RAM (MB) & 2048 & 2048 & 2048 & 512 \\
Swap space (MB) & 4096 & 4096 & 4096 & 1024 \\
Hard disk (GB) & 72 & 72 & 72 & 18 \\
OS (32 bit) & CentOS 5 & Debian 5 & Ubuntu 8.04 LTS & CentOS 5 \\
Application & Apache & PostgreSQL 8.4.1 & Samba & - \\
Benchmark & siege & pgbench & dbench & - \\
Metric & Transact./s & Transact./s & MB/s & - \\
\bottomrule
\end{tabular}
\caption{Load profile and hardware environment for the VM application
benchmarks.}
\label{tab:infrastructure_performance}
\end{table}

The benchmark involves the following scenarios:

\begin{enumerate}
  \item Each application is run individually for one hour on the physical host
  to get reference results.
  \item Each application is run individually for one hour in its virtual machine
  to establish a baseline. The virtual machines baseline results are compared to
  the results of the reference host.
  \item All application workloads are run concurrently for one hour, each in its
  virtual machine. These results are compared to the baselines obtained in the 
  second scenario.
\end{enumerate}

As shown in table \ref{tab:infrastructure_performance}, the results of
the above described tests are compared regarding different metrics.  The
results are presented below.

\begin{figure}
\begin{center}
\subfloat[Web server performance test results.]{
	\label{fig:vserver-infrastructure-perf-ws}
	\includegraphics[scale=0.9]{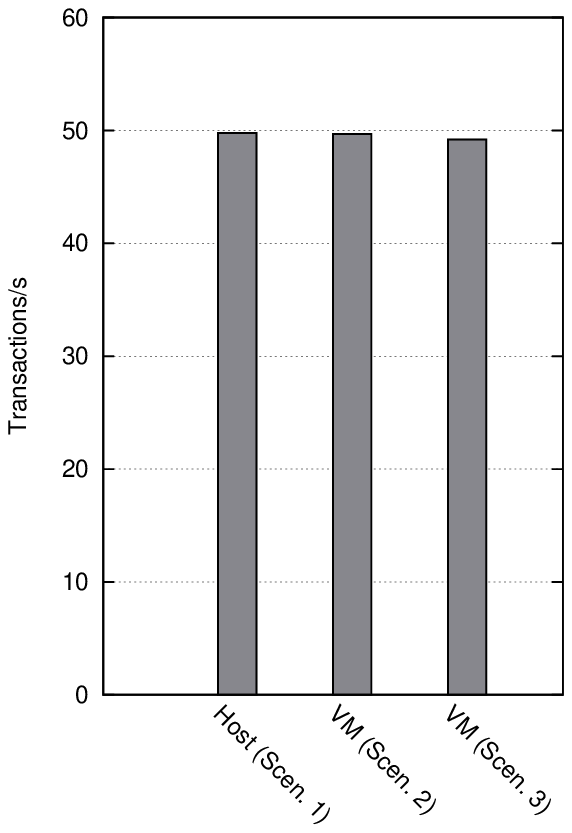}}
%\hspace{0.5cm}
\subfloat[Database server performance test results.]{
	\label{fig:vserver-infrastructure-perf-ds}
	\includegraphics[scale=0.9]{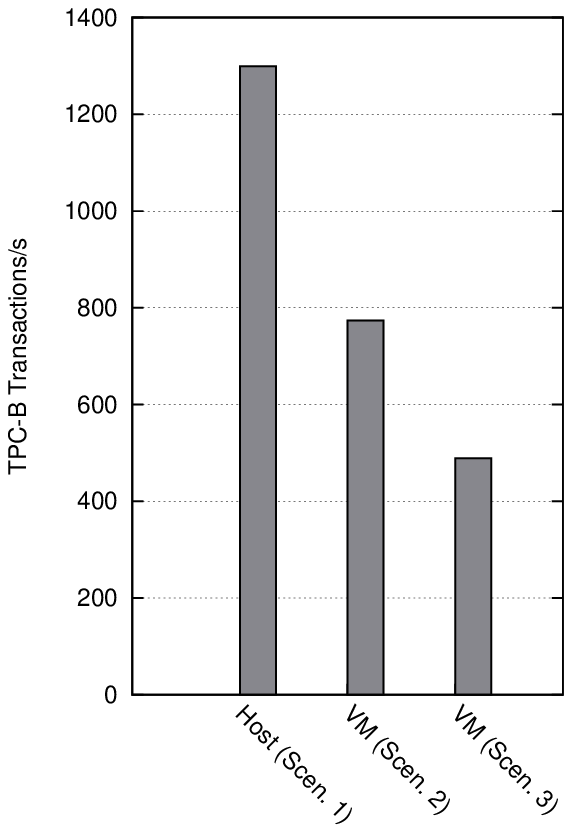}}
\hspace{0.5cm}
\subfloat[File server performance test results.]{
	\label{fig:vserver-infrastructure-perf-fs}
	\includegraphics[scale=0.9]{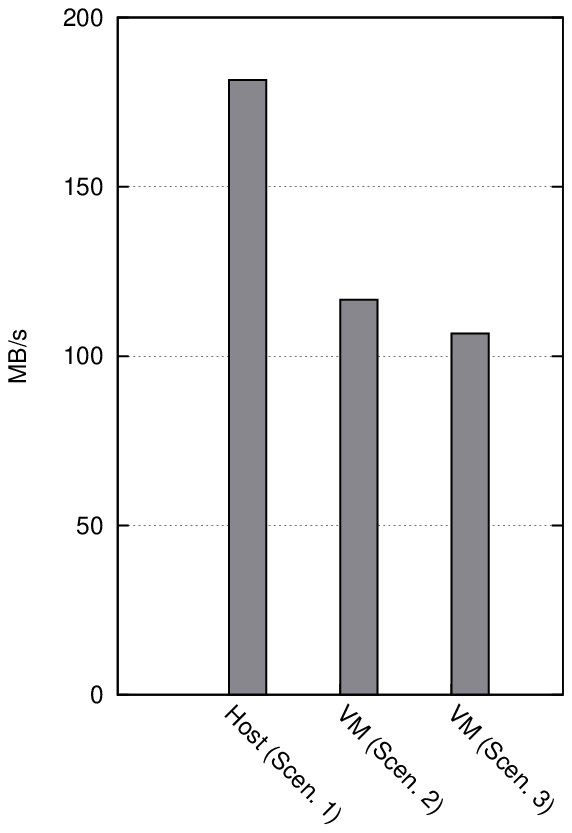}}
\end{center}
\caption{Performance test results of three application programs.}
\label{fig:vserver-infrastructure-perf}
\end{figure}

The web server performance tests show equal results for all three scenarios. 
The non-virtualized Linux system was able to make an average of 49.77
HTTP transactions/s after the one hour individual test run.  The virtual machine
reached 49.68 transactions/s in its individually run scenario.  This is a
performance score of 99.82\% compared to the non-virtualized environment.  In
the third scenario, all virtual machines were simultaneously under load with 
respect to their tested services.  In this scenario the web server performance 
of the virtual machine was only slightly behind and performed 49.20 
transactions/s, which is a performance score of 98.85\%. The results of the 
three scenarios for the web server tests are shown in figure 
\ref{fig:vserver-infrastructure-perf-ws}.

For the database server test, the non-virtualized Linux system performs better
than the virtualized one in their individual runs (scenario 1 vs. scenario 2). 
The non-virtualized Linux was able to process 1299.33 TPC-B\footnote{
Transaction Processing Performance Council (TPC) Benchmark B (TPC-B) measures
throughput in terms of how many transactions per second a system can 
perform \cite{tpc1994}.} transactions/s, whereas the virtual machine processed 
773.75 transactions/s.  Thus, the virtual machine was able to achieve a
performance score of 59.55\%.  In scenario 3, the virtual machine running the
database server performed weaker than in scenario 2: pgbench reported 488.71 
transactions/s, which is a performance score of only 37.61\%.  The results of
the three scenarios for the database server tests are shown in figure 
\ref{fig:vserver-infrastructure-perf-ds}.  The significant performance loss 
between the Scenario 1 and 2 is due to the fact, that a lot of I/O requests 
sent from the host to the NFS backend are waiting.  Thus, the networking 
subsystem can be considered a bottleneck in this scenario.

During the file server tests, the non-virtualized Linux performed better than
the virtual machine in their individual scenarios.  It managed to reach a 
throughput of 181.56 MB/s in its individual run, while the virtual machine 
reached a throughput of 116.67 MB/s.  The virtual machine achieved a performance
score of 64.26\%.  In scenario 3, the throughput of the virtual machine was
slightly weaker (106.70 MB/s), which equals a performance score of 58.77\%.  The
results of the three scenarios for the web server tests are shown in figure 
\ref{fig:vserver-infrastructure-perf-fs}.  The performance gap between the 
non-virtualized scenario and the virtualized ones is also caused by the NFS 
connections to the virtual disk images.

% -----------------------------------------------------------------------------

\section{Conclusion}

In this paper, the network performance of Xen-based virtual machines as well as
the performance of typical application programs for small-to-medium businesses
that may reasonably run in virtual machines have been investigated.  

The benchmark results show, that the Xen hypervisor provides sufficient 
capacities to offer good network performance in terms of latency and throughput
to up to 20 virtual machines on the same host.

The applications performance tests that have been carried out, show that the Xen
hypervisor offers --- with slight constraints --- decent performance for typical
small-to-medium business application environments such as web servers, file 
servers, or database servers.  The performance score in the virtualized
environment was roughly around 75\% of the non-virtualized environment.  This is
an acceptable result, given the fact that the disk images of the virtual
machines are stored on a network shared storage (NFS).

The performance penalties of running these applications at high load in 
multiple, different virtual machines at the same time are around 10\% compared
to running their virtual machines exclusively on the host machine.  The
performance score here is at 65\% of the standalone virtual machine 
environments.

% -----------------------------------------------------------------------------

%\begin{acks}
%\end{acks}

\bibliographystyle{acmtrans}
\bibliography{document}
\begin{received}
%Received February 1986;
%November 1993;
%accepted January 1996
\end{received}

\end{document}